\documentclass[11pt,twoside]{article}

\usepackage{asp2006}
\usepackage{epsf}
\usepackage{psfig}
\usepackage{lscape}

\begin{document}
\title{Planetesimal accretion in binary star systems}   
\author{F. Marzari}   
\affil{Dept. of Physics, University of Padova, Italy}
\author{P. Thebault}
\affil{Observatoire de Paris, Section de Meudon, F-92195 Meudon Principal Cedex,
France and Stockholm Observatory, Albanova Universitetcentrum,
SE-10691 Stockholm, Sweden}    
\author{H. Scholl}
\affil{Observatoire de la C\^ote d'Azur, Nice, France}

\begin{abstract} Numerical simulations of planetesimal accretion 
in circumprimary and circumbinary orbits are described. 
The secular perturbations by the companion star and gas drag 
are included in our models. We derive limits on the 
parameters of the binary system for which accretion and 
then planetary formation are possible. In the circumbinary
case we also outline the radial distance from the baricenter of the
stars beyond which accumulation always occurs. Hydrodynamical 
simulations are also presented to validate our N--body approach 
based on the axisymmetric approximation for the gas of the disk.
\end{abstract}

\section{Formation of planets by core--accretion}

The formation of terrestrial planets and cores of
giant planets within circumstellar disks involves the 
accumulation of a large number of planetesimals, 
solid bodies with initial sizes of roughly several kilometers
(Lissauer 1993; Wetherill \& Stewart 1993).
The initial growth of the planetesimals can follow different paths depending
on the their mutual velocities. If runaway growth occurs, a 
limited number of large planetary embryos form on 
a short timescale (about $10^{4} - 10^5$ years) followed
by a period of violent mutual collisions until 
the planets reach their final mass. If the 
encounter velocities exceeds the planetesimal's escape velocities,
the size distribution of the entire population exhibits an 
orderly growth until larger bodies are formed
on a much longer timescale.

Most observed extrasolar planets are believed to have formed from planetesimals.
The core--accretion model (Pollack et al., 1996) 
seems to explain a large fraction of the observed 
physical and dynamical properties of 
extrasolar gaseous giants
in particular after the inclusion of
migration by interaction with an
evolving disk and gap formation (Alibert et al. (2005)).
Neptune--size extrasolar planets possibly formed 
directly by planetesimal accumulation without reaching the 
critical mass to accrete a massive gaseous envelope. 
Around single stars the efficiency of planetesimal accumulation 
is very high, leading easily to planet formation. 
The influence of collective perturbations like stirring by 
mutual gravitational perturbations and damping by
collisions and gas drag 
effects has been studied in detail in order 
to understand the conditions favoring runaway growth. 
  
Radial velocity surveys have shown that exoplanets are found 
also in binary or higher multiplicity stellar systems 
(Raghavan et al. 2006, Desidera \& Barbieri 2007).
Planetesimal accumulation and then planet formation 
in binary (or multiple) stellar systems appears to be a more 
complex process than around single stars.
The gravitational secular perturbations by the 
companion star may overcome the mutual planetesimal 
interactions and significantly affect the initial stage
of accretion by exciting high eccentricities and then affecting
significantly the relative velocity distribution. 
In this paper we explore the velocity 
evolution of planetesimals in S or C--type orbits under both, the 
perturbing effects of 
the companion star and gas drag. 
We recall that planetesimals revolving just about one star in a binary pair 
are on so-called "S-type" pr "circumprimary" orbits, 
whereas those that revolve about
both stars have "P-type" or "circumbinary" orbits.

\section{The circumprimary case (S--type orbits)}

The size distribution of planetesimals evolves via 
mutual collisions between the bodies. It is crucial that
in the early stages of accumulation the 
relative velocities between the planetesimals remain low. 
While in a planetesimal swarm around a single star 
the relative velocities are on average less than the 
escape velocity of the largest bodies, 
the relative velocities 
can be pumped up to values leading to disruption of the impacting
bodies when a binary
companion is present on an outer orbit. 
Under this condition, fragmentation would dominate over
accretion halting the planetary formation process. 
We have shown (Marzari \& Scholl, 2000) that a crucial role 
is played by gas drag which damps the eccentricity 
forced by the binary companion and causes
an alignment of the planetesimal perihelia. The resulting  
phasing of the orbits leads to very low relative velocities 
between equal size planetesimals. However, once larger 
planetesimals are formed, the perihelion alignment is not
so effective. Different size planetesimals have their 
orbits oriented towards different directions since the 
gas friction depends on the body size (Th\'ebault et al., 2006). 
The forced unpaired
orbital alignment may easily re-establish high random 
velocities thus slowing down or even preventing accretion. 

By performing N--body numerical simulations where the orbits
of the planetesimals are computed under the influence of the 
gravitational pull of the companion star and gas drag, 
we have tested the chances of planetesimal accretion for
120 different binary systems with semimajor axis $a_B$ ranging from
10 to 50 AU and eccentricity from 0.05 to 0.9. The mass ratio between 
the stars have been kept constant and equal to 0.5.  In 
Fig. 1 we plot the regions in the binary parameter space
where accretion is possible via 
runaway growth (dark green) and probably via orderly growth 
(light green). The orange and red areas correspond to scenarios
where planetesimal accretion is inhibited by high 
values of relative velocities while the yellow region is an 
intermediate zone where the tendency towards either accretion or
erosion strongly depends on the planetesimal physical parameters. 

 \setcounter{figure}{0}
 \begin{figure}[!h]
 \plotone{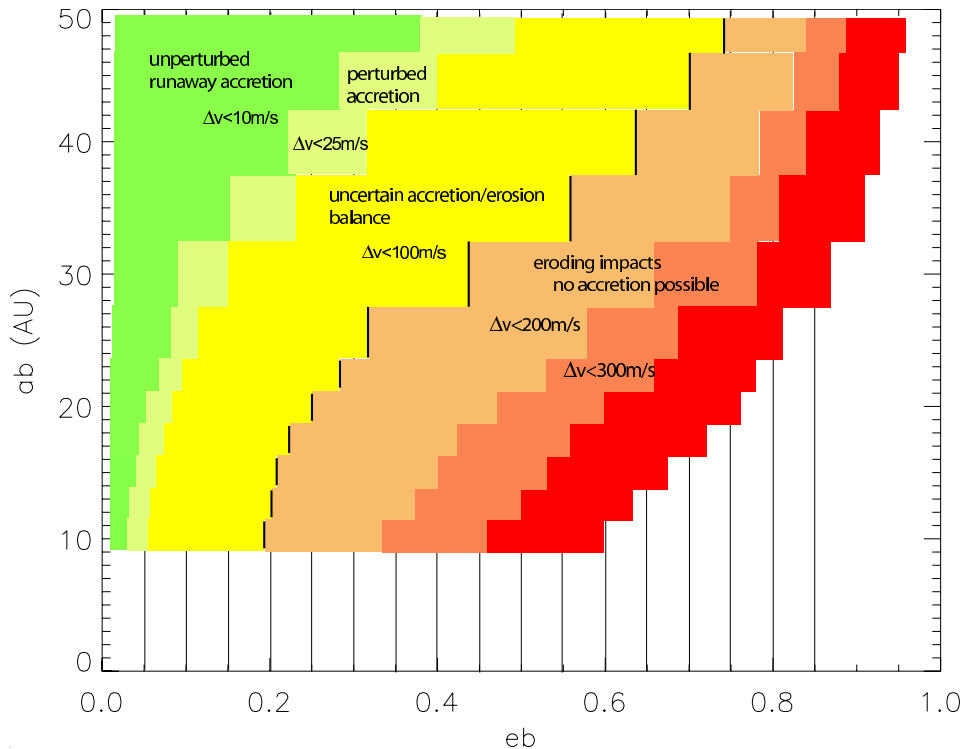}
 \caption{Encounter velocities averaged, over the time interval
$0<t<2\times10^{4}$yrs, between $R_1=2.5$ and $R_2=5\,$km bodies
at 1 AU from the primary star, for different values of the
companion star's semi-major axis and eccentricity.
The short black vertical segments mark the limit beyond which
$<\Delta v_{(R1,R2)}>$ values correspond to eroding impacts
for all tested collision outcome prescriptions.
 }
 \end{figure}

According to Fig. 1 binary star systems with high eccentricity 
and low separation can hardly allow planetesimal accumulation
around the primary star. From the data of the simulations,
Th\'ebault et al. (2006) derived an empirical fit that 
allows to analytically compute the values of binary separation 
$a_B$ and
eccentricity $e_B$ for which accretion is possible: 
\begin{equation}
e_{b} \simeq 0.013 \left(\frac{a_b}{10{\rm AU}}\right)^2
\label{ebab1}
\end{equation}

By extrapolating the fit to larger values of  $a_b$ one can figure out
that for binary separation $a_b \geq \simeq 90$AU the planetesimal
accretion process is not significantly perturbed by the companion
star gravity.

\section{The circumbinary case (P--type orbits)}

So far, only one planet, HD 202206c,  has been found in a P--type orbit  
(Udry et al., 2002; Correia et al. 2005). However, this does not imply 
that circumbinary planets are rare as to detect such planets 
by radial velocity measurements is intrinsically difficult due to the
short--term large--amplitude velocity of the primary induced 
by the companion star. Circumbinary material has been found around 
pre--main--sequence close binaries like DQ Tau or UZ Tau by mid--infrared
surveys. 
The inferred disks are even more massive than the minimum--mass 
solar nebula suggesting that planet formation may undergo in the standard
way. 
As for the circumprimary case, we have explored planetesimal accumulation 
in P--type orbits by performing N--body numerical integrations of 
planetesimals orbits around the barycenter of the binary system. 
In our model we adopt a simplified
approach to compute the gas friction on the bodies by assuming that
the gaseous disk is axisymmetric and pressure supported. Taking into 
account that the tidal force of the binary leads to a gap opening
in the inner disk and to spiral density waves propagating
through the disk we had to test whether the spiral structure 
of the gas density might affect the
planetesimal trajectories altering the orbital alignment due to gas friction
and, in general, changing the planetesimal dynamical evolution.
With a hybrid approach, we have computed the evolution of the 
gaseous component of the disk with an hydrodynamical code
(FARGO, Masset 2000) 
and used  the derived local gas density and velocity 
to calculate the drag force on a limited number of planetesimals
embedded in the disk. The limitation in the number of
computed planetesimal trajectories is related to the
large amount of CPU time required by the hydrodynamical
part of the hybrid code. The outome of a test simulation is given
in Fig. 2. We show the evolution of the orbital eccentricity
and pericenter of 14 planetesimals with a diameter of
10 km and
a density of 2 g/cm$^3$ with equally spaced initial semimajor axes 
between 1.2 and 2.5 AU from the barycenter of the stars. 
The binary system is made of two stars with masses of 0.8 and 0.2 
solar masses, respectively, orbiting each other with a semimajor axis
of 0.2 Au and an eccentricity of 0.4.

 \setcounter{figure}{0}
 \begin{figure}[!h]
 \plottwo{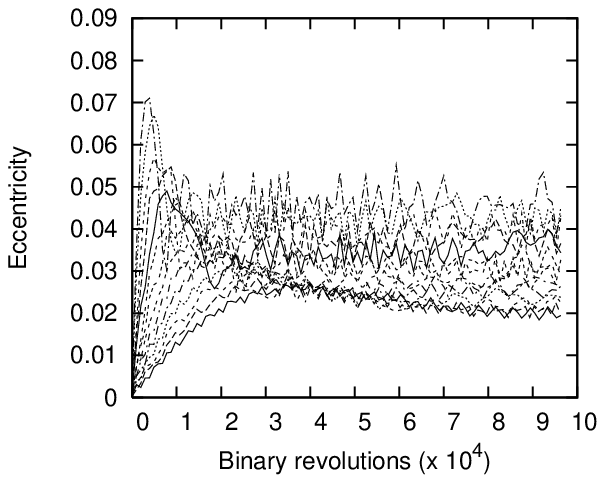}{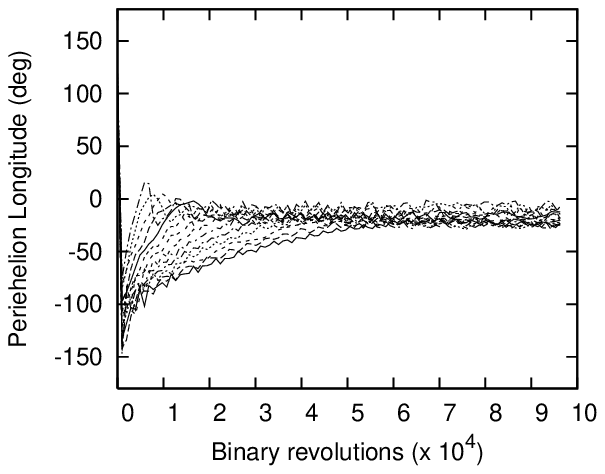}
 \caption{Orbital evolution of 10--km size planetesimals computed
 with the hybrid code. The eccentricities reach a steady state
 while the pericenters are well aligned.
}
\end{figure}

The results of the hybrid code are very similar to those obtained 
with the N--body code where the axisymmetric assumption is adopted. 

After the validation of the N--body code, we have considered a more 
general circumbinary case where the stars are separated by 1 AU 
and have a total mass of 1 $M_{\odot}$.  
We have integrated the trajectories of 
25000 planetesimals with intial semimajor axis ranging from 4 (outside the 
tidal gap) to 12 AU from the baricenter of the two stars
(Scholl et al., 2007). Taking into
account the dependence of the perihelion aligment on the planetesimal size,
even in this case we computed relative velocities 
betwen representative pairs of different size planetesimals. 
In Fig. 3 we display, for different binary mass ratios and 
eccentricities, the minimum radial distance beyond which 
planetesimal accretion and then planetary formation is possible. 

 \setcounter{figure}{0}
 \begin{figure}[!h]
 \plotone{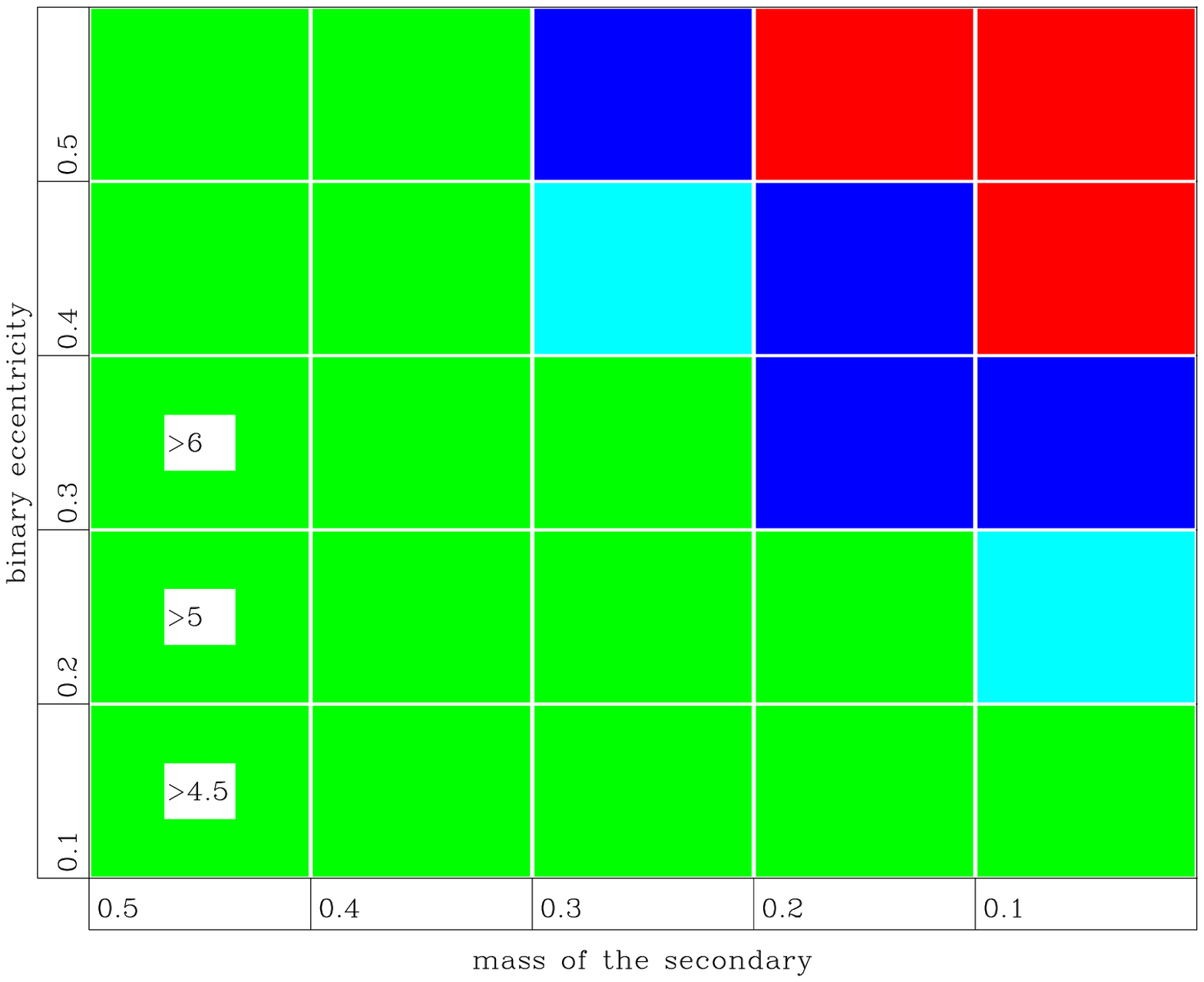}
 \caption{Map of the radial distance $r_l$  beyond which planetary
 formation is possible as a function of the binary mass ratio
 $q = m_2 / (m_1 + m_2) = m_2$ (recall that $m_1 + m_2 = 1 M_{\odot}$)
 and binary eccentricity $e_B$. The color coding is the  following:
 -$Green$: $r_l\leq4$AU (the inner edge of our planetesimal disc)
 -$Pale \quad blue$: 4AU$<r_l\leq 6$AU
 -$Dark \quad blue$: 6AU$\leq r_l \leq 9$AU
 -$Red$: 9AU$\leq r_l<12$AU
 -$Black$: $r_l \geq 12$AU (the outer edge of our planetesimal disc)
 The radial distance given at the center of some rectangles is
 the minimum value beyond which
 $runaway$ accretion
 is possible.  }
\end{figure}

It can be seen that for equal mass stars (mass ratio $q = 0.5$) planet 
formation proceeds in all the
regions of the disk and for all binary eccentricities.
For smaller values of mass ratios and high binary eccentricities,
the inner border for accretion shifts to larger radial
distances. For $q = 0.1$ and $e_b = 0.5$, for example,
planet formation can occur only beyond 10 AU. The strong secular perturbations
due to the large eccentricity and low mass ratio of the binary
prevent planetesimal accumulation closer to the barycenter.
In most cases where accretion is possible, however, the 
growth path is probably not runaway since the perturbations 
of the binary lead to random velocities slightly higher than
the escape velocities from the larger planetesimals. 
 
\section{Conclusions}

Numerical simulations of planetesimal evolution support the 
scenario in which planet formation may undergo even in 
binary star systems. Planetesimals in both S--type and 
P--type orbits keep their relative velocities low enough 
to allow accumulation rather than fragmentation for a wide
range of binary orbital and physical parameters even if orderly 
growth or the so--called Type II runaway growth
\citep{kort01} are possibly more common than the 
conventional fast runaway growth presumed to occur 
around single stars. Both terrestrial planets and 
giant planets are supposed to form in binary systems unless 
extreme orbital conditions for the two stars are met like
large binary eccentricity or very short separation (in the 
case of circumprimary disks). The potential lower rate of 
planet discovery around double stars may be ascribed to these
cases rather than to a general effect related
to the presence of a companion star. 


\acknowledgements 


\end{document}